\documentstyle[12pt,lnfprep,epsfig]{article}

\newcommand{\Header}{
  \begin{tabular}{rl}
  \hspace{-.4cm}\includegraphics{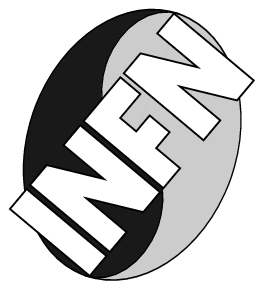} &
    \renewcommand{\arraystretch}{0.5}
    \begin{tabular}{r}
      {\hspace{1cm}~\LARGE\sffamily LABORATORI~ NAZIONALI~ DI~ FRASCATI}\\
      \\
      {\Large\sffamily SIS-Pubblicazioni}\\
    \end{tabular}
    \renewcommand{\arraystretch}{1}
  \end{tabular}
  \vskip 1cm
  \begin{flushright}
  \renewcommand{\arraystretch}{0.5}
    \begin{tabular}{r}
      {\underline{LNF-99/007(P)}}\\      
      {\small 8 Marzo 1999}      \\      
      \\
      {\small\tt hep-ph/9903284} 
    \end{tabular}
  \end{flushright}
  \renewcommand{\arraystretch}{1}
  \vskip 1 cm
  }
\newcommand{\beq}{\begin{equation}}
\newcommand{\eeq}{\end{equation}}
\newcommand{\ba}{\begin{array}}
\newcommand{\ea}{\end{array}}
\newcommand{\beqa}{\begin{eqnarray}}
\newcommand{\eeqa}{\end{eqnarray}}
\newcommand{\lsim}{\stackrel{<}{_\sim}}

\newcommand{\tq}{{\tilde q}}

\newcommand{\tg}{{\tilde g}}

\newcommand{\cA}{{\cal A}}
\newcommand{\cO}{{\cal O}}
\newcommand{\cM}{{\cal M}}
\newcommand{\cP}{{\cal P}_T}
\newcommand{\czdot}{\! \cdot \!}
\newcommand{\no}{\nonumber}
\newcommand{\dfrac}{\displaystyle \frac}
\newcommand{\dint}{\displaystyle \int}

\newcommand{\PL}[3]{{\it Phys.\ Lett.\ }        {\bf #1}, {#3} {(19#2)}}

\newcommand{\PRL}[3]{{\it Phys.\ Rev.\ Lett.\ } {\bf #1}, {#3} {(19#2)}}
\newcommand{\PR}[3]{{\it Phys.\ Rev.\ }         {\bf #1}, {#3} {(19#2)}}
\newcommand{\NP}[3]{{\it Nucl.\ Phys.\ }        {\bf #1}, {#3} {(19#2)}}

\newcommand{\JHEP}[3]{{\it JHEP\ }              {\bf #1}, {#3} {(19#2)}}
\newcommand{\SJNP}[3]{{\it Sov. J. Nucl. Phys. }{\bf #1}, {#3} {(19#2)}}
\newcommand{\YF}[3]{{\it Yad.\ Fiz.\ }          {\bf #1}, {#3} {(19#2)}}
\newcommand{\PTP}[3]{{\it Prog. Theor. Phys. }  {\bf #1}, {#3} {(19#2)}}

\begin{document}
\begin{titlepage}
\title{ 
  \Header
  {\large \bf The $CP$--conserving contribution to the \\ 
  transverse muon polarization in $K^+ \to \mu^+ \nu \gamma$}
}
\author{Gudrun Hiller and Gino Isidori\\
{\it INFN, Laboratori Nazionali di Frascati, P.O. Box 13,
I-00044 Frascati, Italy}
} 
\maketitle
\baselineskip=14pt

\begin{abstract}
We present a detailed estimate of the transverse 
muon polarization ($P_T$) 
due to electromagnetic final--state interactions 
in the decay $K^+ \to \mu^+ \nu \gamma$. 
This $CP$--conserving effect represents the dominant contribution 
to  $P_T$ within the Standard Model. 
As a result of an explicit calculation, we find that
the  $CP$--conserving contribution to $P_T$
is quite small, typically of $\cO(10^{-4})$, 
essentially due to the suppression factor $\alpha/4\pi^2$. 
This enforces the sensitivity of $P_T$ in 
probing extensions of the Standard Model with
new sources of time--reversal violations.
A brief discussion about possible $CP$--violating
contributions to $P_T$ in the framework of
supersymmetric models with unbroken $R$--parity
is also presented.
\end{abstract}

\vspace*{\stretch{2}}
\begin{flushleft}
\end{flushleft}
\begin{center}
Submitted to Physics Letters B
\end{center}
\end{titlepage}
\pagestyle{plain}
\setcounter{page}2
\baselineskip=17pt

\section{Introduction}

The transverse muon polarization in $K^+ \to \pi^0 \mu^+ \nu $    and 
$K^+ \to \mu^+ \nu \gamma$ decays ($P_T$) or the component 
of the muon spin perpendicular to the decay planes, is  
an interesting observable to search for non--standard sources of 
time--reversal violations \cite{Sakurai}. Measurements of $P_T$ 
at the level of $10^{-3}$ are presently undergoing in both channels
at the KEK E246 experiment \cite{Kuno}, similar sensitivities could 
in principle be achieved also at DA$\Phi$NE \cite{Privitera} 
and a proposal to reach $\sigma(P_T)\sim 10^{-4}$ exists at 
BNL \cite{Kuno}.

The transverse muon polarization can be different from zero 
only if the form factors of the corresponding decay amplitude
have non--vanishing relative phases. These can be generated 
either by $CP$--violating couplings 
(assuming $CPT$ conservation)
or by absorptive effects due to electromagnetic 
final--state interactions (FSI). The 
$CP$--violating contribution is
absolutely negligible within the Standard Model (SM)
and therefore if detected at the level of $10^{-4}$
or above it could only be
of non--standard origin. In specific New Physics 
models $CP$--violating contributions to $P_T$
as large as $10^{-3}$ can be generated both 
in  $K^+ \to \pi^0 \mu^+ \nu $ (see e.g. \cite{Vari1,WuNgold,Vissani})
and in  $K^+ \to \mu^+ \nu \gamma$  (see e.g. \cite{Vari3,WuNg,CGL}).
As pointed out by Wu and Ng \cite{WuNg}, 
the two modes are complementary in testing possible 
New Physics scenarios. 

Electromagnetic FSI effects in $K^+ \to \pi^0 \mu^+ \nu$ 
appear only at the two--loop level, leading to 
a $CP$--conserving contribution to $P_T$  
which is $\cO(10^{-6})$ and thus safely negligible \cite{Zit}. 
The situation is however different in 
$K^+ \to \mu^+ \nu \gamma$, where FSI effects are present 
already at the one--loop level and could potentially
lead to $\cO(10^{-3})$ effects, like in the case of 
$K_L \to \pi^- \mu^+ \nu$ \cite{OK}. In principle 
$CP$--conserving and $CP$--violating contributions to   $P_T$ 
could be disentangled by a measurement of this quantity
in both $K^+$ and $K^-$  $CP\mbox{--conjugated}$ modes \cite{OK}, 
however at present this is not experimentally feasible. 
A detailed theoretical estimate of the 
FSI contributions to $P_T(K^+ \to \mu^+ \nu \gamma)$,
which is the main purpose of the present letter,
is therefore needed in order to identify possible 
New Physics effects.

The paper is organized as follows: in section 2 we introduce 
observables and  kinematics of $K^+ \to \mu^+ \nu \gamma$,
we then proceed in section 3 
estimating the dominant FSI contribution 
to $P_T$, induced by $\pi^0$ exchange. 
A brief discussion about possible supersymmetric 
contributions to $P_T$ is presented in section 4 and
the results are summarized in the conclusions.

\section{Observables and kinematics}
Assuming purely left--handed neutrinos,
the matrix element of the transition
$K^+(p) \to \mu^+(l) \nu (p_\nu) \gamma(q,\epsilon)$
can be generally decomposed as
\beq
\cM = - \frac{i e G_F}{\sqrt{2}} \sin\theta_c ~ \epsilon^*_\alpha~
\bar{u}(p_\nu) (1+\gamma_5) \left[  T_1^\alpha
+ \gamma_\beta  T_2^{\alpha\beta} + i \sigma_{\beta\gamma} 
T_3^{\alpha\beta\gamma} \right] v(l) ~,
\label{eq:MM}
\eeq
where we have factored--out explicitly the dependence 
on the Fermi constant ($G_F$) and the Cabibbo angle ($\theta_c$).
Within the SM, the  tensors $T_i$ are given by
\beqa
T_1^\alpha &=& m_\mu f_K \left( \frac{p^\alpha}{q\czdot p}
- \frac{l^\alpha}{q\czdot l} \right)~, \label{T1} \\
T_2^{\alpha\beta} &=& \frac{F_A}{m_K} \left( g^{\alpha\beta} {q\czdot p}  
-p^\alpha q^\beta\right) - i \frac{F_V}{m_K}
\epsilon^{\alpha\beta\rho\delta} q_\rho p_\delta~,\\
T_3^{\alpha\beta\gamma} &=& m_\mu f_K 
\frac{g^{\alpha\gamma} q^\beta}{2 q\czdot l}~, 
\label{T3}
\eeqa
where $f_K \simeq 160$ MeV is the usual kaon decay constant and 
$F_{V,A}$  are form factors associated with the time--ordered 
product of electromagnetic and weak currents \cite{bijnens}.
The one--loop estimate of $F_{V,A}$ in chiral
perturbation theory leads to the constant values
$F_V=-0.095$ and $F_A=-0.042$, which are consistent 
with experimental data \cite{bijnens}.
As long as electromagnetic FSI and 
$\cO(G_F^2)$ electroweak corrections are neglected,
$f_K$ and $F_{V,A}$ are real. On the other hand,  
new sources of $CP$ violation beyond the 
Cabibbo--Kobayashi--Maskawa (CKM)  matrix \cite{CKM}
could generate non--negligible phases to these 
form factors \cite{WuNg}.

Introducing the adimensional variables
\beq
x=\frac{2 p\czdot q}{m_K^2}~, \qquad 
y=\frac{2 p\czdot l}{m_K^2} \qquad\mbox{and}\qquad
r_{\mu}=\frac{m^2_{\mu}}{m_K^2}~,
\eeq
the Dalitz plot distribution of 
the decay can be written as
\beq
\frac{\mbox{d}^2\Gamma}{\mbox{d}x \mbox{d}y} = 
\frac{ m_K }{256 \pi^3} \sum_{pol(\nu,\gamma,\mu)} \left| \cM
\right|^2
= \frac{ G_F^2 m_K^5 \alpha \sin^2 \theta_c  }{ 32 \pi^2 } \rho(x,y)~,
\eeq
with the boundary conditions 
\beqa
0 \leq &x& \leq 1-r_{\mu}~, \no \\
\frac{r_{\mu}+(1-x)^2}{1-x} \leq &y& \leq 1 + r_{\mu}~.
\eeqa
The full expression of the adimensional function $\rho(x,y)$, 
calculated with the $T_i$ in (\ref{T1}~-~\ref{T3}),  
can be found in \cite{WuNg}. The inner bremsstrahlung 
(IB) contribution, 
obtained in the limit $F_V=F_A=0$, is given by
\beqa
&& \rho_{_{F_V=F_A=0}}(x,y)  \doteq  \rho_{IB}(x,y) 
   = 2 r_{\mu} \frac{f^2_K}{m^2_K} 
   \left( \frac{1-y+r_{\mu}}{x^2(x+y-1-r_{\mu})} \right) 
   \no \\
&&\qquad\qquad\qquad \times \left( x^2+ 2(1-x)(1-r_{\mu}) 
     - \frac{2xr_{\mu}(1-r_{\mu})}{x+y-1-r_{\mu}} 
   \right)~.
\label{rhoIB}
\eeqa

The triple product defining the 
transverse muon polarization in $K^+ \to \mu^+ \nu \gamma$ is 
\beq
\cP \doteq \frac{\vec{s} \czdot (\vec{q} \times \vec{l} )}
{|\vec{q} \times \vec{l} |}~,
\eeq
where $\vec{q}$, $\vec{l}$ and $\vec{s}$ 
denote the spatial components in the kaon rest frame of
$q$, $l$ and the spin vector of the muon $s$,
respectively. As usual, the latter  
obeys the relations $s\czdot l=0$ and $s^2=-1$. 
In terms of four--dimensional vectors we can write
\beq
 \cP = \frac{2}{m_K^3} 
\frac{\epsilon_{\alpha\beta\gamma\delta} p^\alpha s^\beta q^\gamma
l^\delta}{F(x,y)}~
\qquad\quad (\epsilon_{0123}=+1)~,
\eeq
where
\beq
F(x,y) = 
\sqrt{(1-y+r_{\mu})\left[(1-x)(x+y-1)-r_{\mu}\right]}~.
\eeq
The modulus squared of $\cM(K^+ \to \mu^+ \nu \gamma)$, 
summed over photon and neutrino polarizations, can be 
decomposed as 
\beq
\sum_{pol(\nu,\gamma)} \left| \cM \right|^2 = 
e^2 G^2_F m_K^4 \sin^2\theta_c \left[ \rho(x,y) +
\cP \sigma(x,y) +\cO(s\czdot p,s\czdot q ) \right]~,
\eeq
therefore the ratio $\sigma(x,y)/\rho(x,y)$ is usually
referred as the distribution of the 
transverse muon polarization \cite{Vari3,WuNg}.
We note, however, that this notation is 
misleading since $\sigma(x,y)/\rho(x,y)$ 
is not directly an observable.
What can be measured in realistic experiments is, for instance,
the expectation value of $\cP$ averaged over a Dalitz plot 
region $S$, which is given by
\beq
\langle P_T \rangle_S \doteq \dfrac{ \dint_S \mbox{d}\Phi 
 \sum_{pol(\nu,\gamma)} \left| \cM \right|^2 ~ \cP }
{\dint_S \mbox{d}\Phi \sum_{pol(\nu,\gamma)} \left| \cM \right|^2
  }~=~
 \dfrac{ \dint_S \mbox{d}x \mbox{d}y~ 
 \sigma(x,y) G(x,y) }{\dint_S \mbox{d}x \mbox{d}y~ \rho(x,y) }~,
\label{PTS}
\eeq
where\footnote{~$\mbox{d}\Phi$ indicates the differential
phase--space element and the second identity in (\protect\ref{PTS}) 
follows from integration over angular variables.}
\beq
G(y) = \frac{1}{2 \beta^2} \left[1+ \frac{\beta^2-1}{2 \beta} 
\log{\left(\frac{1+\beta}{1-\beta}\right)} \right]
\qquad\mbox{and}\qquad
\beta  = \sqrt{1-4 r_{\mu}/y^2}~.
\eeq
In the limit where the area $S$ reduces to a point in the Dalitz plot,
the integrals on the r.h.s. of (\ref{PTS}) drop out and we obtain 
\beq
 P_T (x,y) ~ = ~
 \dfrac{\sigma(x,y) G(x,y) }{\rho(x,y) }~.
\label{PTxy}
\eeq

As it can be easily verified, $\sigma(x,y)$ is non vanishing 
only if the $T_i$ have non--trivial phases. Assuming that
New Physics contributions generate complex $CP$--violating
phases for the form factors $f_K$ and $F_{V,A}$ of 
(\ref{T1}~-~\ref{T3}), we obtain
\beq
\sigma_{NP}(x,y) = - 2\sqrt{r_{\mu}}F(x,y)
\left[ \Im\left(\frac{f_K F_V^*}{m_K}\right) f_V(x,y) +
\Im\left(\frac{f_K F_A^*}{m_K}\right) f_A(x,y) \right]~,
\label{eq:WN}
\eeq
where
\beqa
f_V(x,y) &=& \frac{2-x-y}{x+y-1-r_\mu}~,\no \\
f_A(x,y) &=& \frac{(2-x)(x+y)-2(1+r_\mu)}{x(x+y-1-r_\mu)}~,
\eeqa
in agreement with the results of \cite{Vari3,WuNg}.

\section{The $CP$--conserving contribution to $P_T$}
\begin{figure}[t]
\vskip -1.5truein
\centerline{\epsfysize=8.5in
{\epsffile{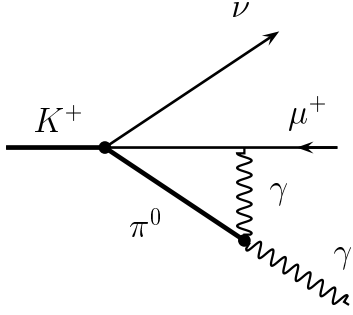}}}
\vskip -4.9truein
\caption[]{The leading rescattering diagram contributing to 
the transverse muon polarization in $K^+ \to \mu^+ \nu \gamma$~.}
\label{fig:piloop}
\end{figure}

As anticipated in the introduction, 
the $CP$--conserving contribution to $P_T$
is generated by  electromagnetic FSI
which induce non--trivial phases in the $T_i$.
The leading rescattering diagram
contributing in $K^+ \to \mu^+ \nu \gamma$
is the one  shown in Fig.~\ref{fig:piloop}. 
Other absorptive 
contributions appearing only at the two--loop 
level can be safely neglected \cite{Zit}.

Since we are interested only in the absorptive contribution 
of the diagram in Fig.~\ref{fig:piloop}, 
the result is finite and can be calculated in terms of
on--shell amplitudes for $\pi^0\to\gamma\gamma$ and
$K^+\to\pi^0\mu^+\nu$. In our conventions these are given by
\beqa
\cA\left( \pi^0(p_\pi)\to\gamma(\epsilon_1,q_1)\gamma(\epsilon_2,q_2)
 \right) &=& +
 \frac{i e^2 }{2\sqrt{2} \pi^2 f_\pi} \epsilon_{\mu\nu\rho\sigma}
\epsilon_1^\mu q_1^\nu \epsilon_2^\rho q_2^\sigma~, \\
\cA\left( K^+(p)\to\pi^0(p_\pi)\mu^+(l)\nu(p_\nu) \right) &=& -
\frac{i G_F}{2} \sin\theta_c ~\left[ f_+(p+p_\pi)^\mu
+ f_-(p-p_\pi)^\mu \right]\qquad  \no \\
&& \qquad\qquad\qquad \times \bar{u}(p_\nu) (1+\gamma_5) \gamma_\mu
v(l) ~.
\eeqa
The explicit analytical expression 
of $\cM_{FSI}$ thus obtained is rather complicated 
and will not be given here,
but it simplifies considerably in the limit
$m^2_{\pi^0}/m_K^2\to 0$. Considering the interference
of $\cM_{FSI}(m_{\pi^0}=0)$ with the leading matrix element
in (\ref{eq:MM}) leads to
\beq
\sigma_{FSI}(x,y) =  \frac{\alpha}{4 \pi^2}~ 
 \sqrt{r_{\mu}}F(x,y)~\frac{f_K}{f_\pi}
 \left[ f_+ \sigma_1+ f_{-} \sigma_2+
\frac{m_K(F_A-F_V)}{f_K} f_{-} \sigma_3 \right]~,
\label{sFSI}
\eeq
where
\beqa
\sigma_1 &=&\frac{1}{4 x (1-x-y)^2} (4+r_{\mu}-4x-4y)
(1+r_{\mu}-2x +x^2-y+x y)~, \\
\sigma_2&=&\frac{r_{\mu}}{4 x (1-x-y)^2} 
(1+r_{\mu}-2 r_{\mu} (x+y)-x^2-3 y + x y +2 y^2)~, \\
\sigma_3&=&\frac{1}{4 (1-x-y)} (1+r_{\mu}-y)(x+y-1-r_{\mu})~.
\eeqa
\begin{figure}
\centerline{\epsfysize=2.6in
{\epsffile{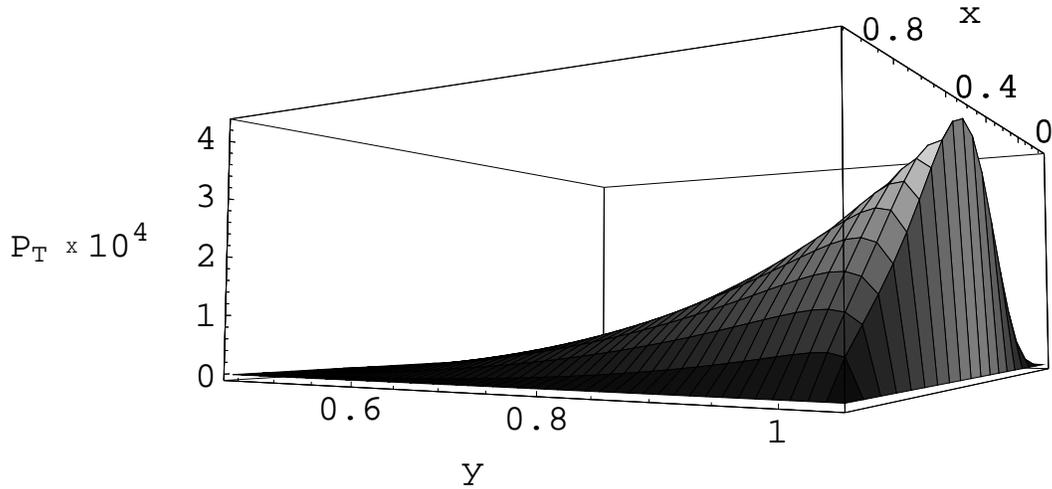}}}
\caption{Distribution of the transverse muon polarization,
as defined in (\ref{PTxy}), resulting from FSI and evaluated 
in the limit $m_{\pi^0}=0$.}
\label{fig:pt0}
\end{figure}  

\begin{figure}
\centerline{\epsfysize=2.6in
{\epsffile{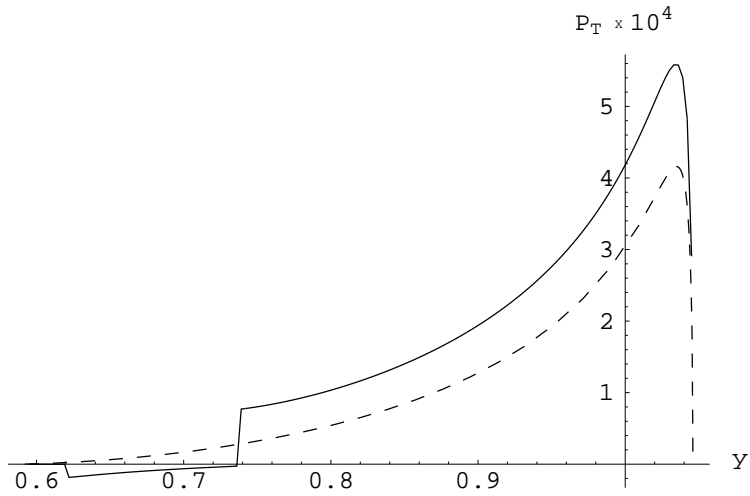}}}
\caption{Distribution of the transverse muon polarization
resulting from FSI for fixed $x=0.5$ as a function of $y$. 
The solid curve correspond the the complete calculation,
the dashed one is obtained in the limit  $m_{\pi^0}=0$.}
\label{fig:pt05}
\end{figure}

The value of $P_T$ obtained inserting 
$\sigma_{FSI}(x,y)$ into (\ref{PTxy}) is 
plotted in Fig.~\ref{fig:pt0} as a function of $x$ and $y$.
A comparison of the full  result obtained  
with $m_\pi^0\not=0$ and the approximate one 
for $m_\pi^0=0$ is shown in Fig.~\ref{fig:pt05}.
Both figures have been obtained using 
$f_+=1$, $f_{-}=0$ and $f_K/f_{\pi}=1.21$, in addition to the 
numerical values of $F_{A,V}$ specified in section 2. 
It is worthwhile to mention that the analytical result 
in (\ref{sFSI}) is very general and could be easily extended
to the case of other radiative semileptonic processes,
like for instance the $B^+\to \tau^+\nu\gamma$ decay.

As it can be read off from Fig.~\ref{fig:pt0}, 
the FSI contribution to $P_T(K^+\to \mu^+\nu\gamma)$ is of 
$\cO(10^{-4})$ all over the phase space. This result,
mainly due to the strong suppression factor $\alpha/4\pi^2$ 
in (\ref{sFSI}), implies that an experimental 
evidence of $P_T$ at the level of $10^{-3}$ would be a clear 
signal of physics beyond the SM. Moreover, we note 
that the peak of $(P_T)_{FSI}$ near the 
boundary $y=1+r_{\mu}$ is simply due to the vanishing 
of the bremsstrahlung contribution in $\rho(x,y)$
(see Eq.~(\ref{rhoIB})): if both $F_{A}$ and $F_{V}$ were 
vanishing, $(P_T)_{FSI}$ would have an unphysical divergence 
(occurring in a region with almost no events) 
for $y\to 1+r_{\mu}$.

\section{Supersymmetric $CP$--violating contributions to $P_T$}
Supersymmetric extensions of the SM with unbroken $R$--parity,
minimal particle content and generic flavour
couplings represent a very attractive possibility 
from the theoretical point of view. In this context several
new $CP$--violating phases are introduced (see e.g. \cite{HKR}
and references therein), therefore sizable $T$--violating 
contributions to $P_T$ could in principle be expected. 
The question we shall try to answer in the following is 
what is the maximum value of these possible $T$--violating
effects.

Supersymmetric contributions to the transverse muon polarization 
in $K^+ \to \mu^+ \nu \gamma$ have been extensively discussed 
by Wu and Ng \cite{WuNg}.
According to these authors, the possibly largest
contribution results from the effective $W\bar{s}_Ru_R$ 
vertex induced  by squark--gluino penguins.
In the framework of the mass--insertion approximation
and employing the so--called super CKM basis \cite{HKR},
the leading contribution to the effective 
$W\bar{s}_Ru_R$ vertex is generated at the second order
in the mass expansion  by  double $\tq^i_L - \tq^j_R$ mixing.
Indeed, in this framework the result of \cite{WuNg} 
can be re--written as
\beq
\Im\left(\frac{f_K F_V^*}{m_K}\right)_{W\bar{s}_Ru_R}
 = -\frac{\alpha_s(M_\tq)}{18\pi\sin\theta_c}
\left[ x^2 I_0(x) \right] \Im\left[ \left(\delta^{D}_{LR}\right)_{23}
\left(\delta^{U}_{LR}\right)_{13}^*\right]~,
\label{dm}
\eeq
where $M_\tq$ $(M_\tg)$ indicates the average squark (gluino) mass,
$x={M^2_\tq / M^2_\tg}$~, $I_0(x)$ is defined in 
\cite{WuNg}\footnote{~$I_0(1)=1$ and 
$[x^2I_0(x)]\sim 1$ for reasonable values of $x$.}
and, denoting by $M^2_{[U,D]}$ the squark mass matrices,
\beq
\left(\delta^{[U,D]}_{LR}\right)_{ab}=
\left(M^2_{[U,D]}\right)_{a_R b_L}/M^2_{\tq}~.
\eeq
Interestingly, the double--mixing mechanism leading to (\ref{dm})
is very similar to the one occurring in the chargino--squark 
contribution to the effective $Z\bar{s}_Ld_L$ vertex, 
relevant for rare decays and recently discussed in \cite{CI}. 
In both cases the supersymmetric contribution is potentially 
large since:
i) there is no explicit $1/M_\tq$ suppression in the 
amplitudes, ii) there are no suppressed CKM matrix elements
and iii) the third generation of squarks is involved. 
However, the  charged--current nature of the  $W\bar{s}_Ru_R$ 
amplitude implies that one of the two 
$\tq^i_L - \tq^j_R$ mixing terms must be in the down sector.
This makes a substantial difference among $W\bar{s}_Ru_R$ and  
$Z\bar{s}_Ld_L$ amplitudes, indeed the coupling
$(\delta^{D}_{LR})_{23}$ appearing in (\ref{dm}) 
is much more constrained than the corresponding one 
relevant for the $Z\bar{s}_Ld_L$ vertex, namely
$(\delta^{U}_{LR})_{23}$. Vacuum--stability 
bounds \cite{CD} and phenomenological constraints 
from the $b\to s\gamma$ decay \cite{GGMS} naturally leads to 
\beq
\left| \left(\delta^{D}_{LR}\right)_{23}
\left(\delta^{U}_{LR}\right)_{13}^* \right| \lsim O(10^{-2})~. 
\eeq
Therefore, unless extremely 
fine--tuned mechanisms are invoked, we estimate 
the r.h.s. of (\ref{dm}) to be at most of
$\cO(10^{-4})$. Using Eq.~(\ref{eq:WN}) we then
obtain an upper bound for the contribution
of the  gluino--mediated $W\bar{s}_Ru_R$ vertex
to $|P_T (K^+ \to \mu^+ \nu \gamma)|$ of about $10^{-4}$, 
that is more than two orders of magnitude smaller than 
reported in \cite{WuNg}.

Similar comments apply to the other mechanism discussed 
in \cite{WuNg}, namely the effective $H\bar{s}_Lu_R$ vertex.
Also in this case the constraints imposed by the
$b\to s\gamma$ transition put severe fine--tuning problems. We then
conclude that in the framework of 
minimal $R$--parity conserving supersymmetric models
it is very unnatural to generate contributions 
to $P_T(K^+ \to \mu^+ \nu\gamma)$ larger than $10^{-4}$.
Analogous conclusions have already been obtained 
by Fabbrichesi and Vissani in the case of 
$K^+ \to \pi^0 \mu^+ \nu$ \cite{Vissani}.

\section{Conclusions}
We have presented a detailed estimate of the transverse 
muon polarization due to electromagnetic final--state interactions 
in the decay $K^+ \to \mu^+ \nu \gamma$. Our analysis
shows that $(P_T)_{FSI}$ is of $\cO(10^{-4})$ all over the 
phase space and that the largest values are obtained 
in the less populated region of the Dalitz plot. 
These results imply that an experimental evidence 
of $P_T$ at the level of $10^{-3}$, or above, would be
a clear signal of physics beyond the SM.

We have further argued why it is very difficult to 
accommodate values of $|P_T(K^+ \to \mu^+ \nu\gamma)|$
larger than $10^{-4}$ in the framework of 
minimal $R$--parity conserving supersymmetric models.
If a non--vanishing value of $P_T(K^+ \to \mu^+ \nu\gamma)$ 
were observed in the forthcoming experiments, 
this could be more likely understood in the framework of
$R$--parity breaking models \cite{CGL} or models with 
a non--standard Higgs sector \cite{Vari3}.

\section*{Acknowledgements}
We thank JoAnne Hewett and Yoshi Kuno for interesting 
discussions that stimulated us to start this work. 
This project is partially supported by the EEC-TMR Program, 
Contract N.~FMRX-CT98-0169.

\end{document}